\documentstyle[preprint, floats, aps]{revtex}

\begin{document}

\preprint{CALT-68-2066, QUIC-96-001}

\draft

\title{Pasting quantum codes}
\author{Daniel Gottesman\thanks{gottesma@theory.caltech.edu}}
\address{California Institute of Technology, Pasadena, CA 91125}
\maketitle

\begin{abstract}
I describe a method for pasting together certain quantum error-correcting 
codes that correct one error to make a single larger one-error quantum 
code.  I show how to construct codes encoding 7 qubits in 13 qubits using 
the method, as well as 15 qubits in 21 qubits and all the other ``perfect'' 
codes.
\end{abstract}


Quantum computers have a great deal of promise, but they are likely to be 
inherently much noisier than classical computers.  One approach to dealing 
with noise and decoherence in quantum computers and quantum 
communications is to encode the data using a quantum error-correcting 
code.  A number of such codes and classes of codes are known
\cite{shor1,calderbank1,steane1,laflamme,bennett,gottesman,calderbank2,steane2}.
However, the only known method of automatically 
generating such codes is to find a suitable classical error-correcting code 
and convert it into a quantum code~\cite{calderbank1,steane1}.  This 
method is limited to producing less efficient codes (i.e., with smaller ratio 
of encoded qubits to total qubits) than dedicated quantum codes, so a method 
of automatically producing highly efficient quantum codes is desirable.  I 
will present here a method to create one-error quantum codes from 
smaller ones with almost no effort.

The conditions for a set of $n$-qubit states $|\psi_1 \rangle, \ldots, 
|\psi_{2^k} \rangle$ to form an error-correcting code for the errors $E_a$ is
\begin{equation}
\langle \psi_i | E_a^\dagger E_b | \psi_j \rangle = C_{ab} \delta_{ij},
\label{conditions}
\end{equation}
where $C_{ab}$ is independent of $i$ and $j$~\cite{bennett,knill}.  A code 
with $2^k$ states encodes $k$ qubits.  Typically, a code will be designed to 
correct all possible errors affecting less than or equal to $t$ qubits.  The 
basis errors $E_a$ are usually tensor products of
\begin{equation}
I = \left( \begin{array}{cc} 1 & 0 \\ 0 & 1 \end{array} \right),\ 
X_i = \left( \begin{array}{cc} 0 & 1 \\ 1 & 0 \end{array} \right),\ 
Y_i = \left( \begin{array}{cc} 0 & -1 \\ 1 & 0 \end{array} \right),\ 
Z_i = \left( \begin{array}{cc} 1 & 0 \\ 0 & -1 \end{array} \right),
\end{equation}
where the subscript $i$ refers to the qubit which the error acts on.

If the matrix $C_{ab}$ has maximum rank, the code is called a {\em 
nondegenerate} code.  If $C_{ab}$ has determinant 0, it is a {\em 
degenerate} code.  Most known codes are nondegenerate codes (in fact, for 
most known codes, $C_{ab} = \delta_{ab}$).  For a nondegenerate code, 
each error acting on each code word must produce a linearly independent 
state.  In order to have enough room in the Hilbert space for all of these 
states, there is a maximum possible efficiency for the code, known as the 
quantum Hamming bound~\cite{ekert}.  For codes to correct one error, the 
quantum Hamming bound takes the form
\begin{equation}
(3n+1) 2^k \leq 2^n.
\end{equation}
If equality holds, the code is known as a {\em perfect} code.  For a perfect 
code, $3n+1$ must be a power of 2.  Since $4^j-1$ is divisible by 3, while 
$2^{2j+1}-1$ is not, there are possible one-error perfect codes for $n=(4^j - 
1)/3$.  The perfect codes have $n-k = 2j$.  The two smallest such codes are 
for $n=5, 21$, but there is a full infinite class of them.  Multiple-error 
perfect codes are much rarer.  There are known $n=5$ 
codes~\cite{laflamme,bennett}, but until now, it was unknown if the other 
perfect codes existed.

Finding a set of states that satisfies condition~(\ref{conditions}) without 
guidance is difficult at best.  In~\cite{gottesman} and~\cite{calderbank2}, 
more powerful group theoretic methods are presented that reduce the task to an 
admittedly still difficult combinatorial problem.  Using the terminology 
of~\cite{gottesman}, a quantum error-correcting code is defined in terms of 
its {\em stabilizer} ${\cal H}$, which is the set of operators $M$ formed 
from products of $X_i$, $Y_i$, and $Z_i$ that fix all of the states in the 
coding space $T$.  $T$ forms the joint $+1$-eigenspace of the operators in 
${\cal H}$.  In order for this to be non-empty, the elements of ${\cal H}$ 
must all commute with each other and square to $+1$.  If ${\cal H}$ is 
generated by $a$~elements, the code encodes $n-a$~qubits.

If an error $E$ anticommutes with $M \in {\cal H}$, when $E$ acts on a 
state $|\psi \rangle$ in $T$, it will take it from the $+1$-eigenspace of $M$ 
to the $-1$-eigenspace, where we can recognize it as an incorrect state, and 
hopefully correct it.  Since all products of $X_i$, $Y_i$, and $Z_i$ commute 
or anticommute, we can define functions $f_M$ and $f$:
\begin{equation}
f_M (E) = \left\{ \begin{array}{ll} 0 & \mbox{if $[M,E]=0$} \\ 1 & \mbox{if 
$\{M,E\}=0$} \end{array} \right.
\end{equation}
\begin{equation}
f (E) = \left(f_{M_1}(E), f_{M_2}(E), \ldots f_{M_a} (E) \right),
\end{equation}
where $M_1, \ldots, M_a$ are the generators of ${\cal H}$.  Given two 
errors $E$ and $F$, if $f(E) \neq f(F)$, then $E |\psi \rangle$ and $F |\psi 
\rangle$ are in different eigenspaces for some element of ${\cal H}$, so 
they are orthogonal, and we can distinguish them and correct them.  
Conversely, if $f(E) = f(F)$, then we cannot properly distinguish $E$ and 
$F$, which will cause a problem unless $E |\psi \rangle$ is actually equal to 
$F |\psi \rangle$, giving us a degenerate code.  In this case, $F^\dagger E 
|\psi \rangle = |\psi \rangle$, so $F^\dagger E \in {\cal H}$.  For a 
nondegenerate code, all of the values $f(E)$ must therefore be distinct, 
allowing $f(E)$ to serve as the error syndrome.  Note that $f(I)=0$, so 
$f(E)$ must be nonzero for nontrivial $E$.

In~\cite{gottesman}, I gave a construction for one-error codes with $n=2^j$, 
$k=n-j-2$.  For all of these codes, the first two generators have the form 
$M_1 = X_1 \ldots X_n$ and $M_2 = Z_1 \ldots Z_n$.  Therefore,
all of the error syndromes for these codes start with $01$ for 
an $X_i$ error, with $10$ for a $Z_i$ error, and $11$ for a $Y_i$ error.  
None of the error syndromes beginning with $00$ are used.  Therefore, we 
can add more qubits and thus more possible errors to the code, so long 
as all the error syndromes for the new errors begin with 
$00$.  The new error syndromes will all have to be different, of course, 
which will necessitate extending most of the generators of ${\cal H}$ to 
have nontrivial action on the new qubits.  

We want the new errors to have $f_{M_1} (E) = f_{M_2} (E) = 0$, so 
we will leave $M_1$ and $M_2$ alone, letting them act trivially 
on the new qubits.  If we extend the remaining generators by pasting on 
the generators of a nondegenerate code with two fewer generators, all of 
the new error syndromes are guaranteed to be distinct, since the smaller 
code must distinguish them to be a good code.  See figure~\ref{scheme} for 
a schematic picture of this process.
\begin{figure}
\begin{picture}(200,100)
\put(0,0){\framebox(140,70){1st code}}
\put(140,0){\framebox(60,70){2nd code}}
\put(140,70){\framebox(60,30){I}}
\put(0,70){\framebox(140,30)}
\put(0,85){\makebox(140,15){XXXXXXXXXXXX}}
\put(0,70){\makebox(140,15){Z\,Z\,Z\,Z\,Z\,Z\,Z\,Z\,Z\,Z\,Z\,Z}}
\end{picture}
\caption{Pasting together the generators for two codes}
\label{scheme}
\end{figure}

Just distinguishing all errors is not sufficient for ${\cal H}$ to define a 
code.  It must also be Abelian and all elements must square to 1.  However, 
each new generator $M=NP$, where $N$ and $P$ are generators from 
existing codes.  They must individually square to 1, so the product also 
squares to 1.  Similarly, another generator $M' = N' P'$ commutes with 
$M$:  $N N' = N' N$ and $P P' = P' P$.  The $N$s and $P$s act on 
different qubits, and therefore commute.  Thus,
\begin{equation}
M M' = (N P) (N' P') = (N' P') (N P) = M' M.
\end{equation}
Therefore, ${\cal H}$ formed by this method will always form a new 
error-correcting code.

The smallest code we can create this way from existing codes is given by 
pasting a 5-qubit code~\cite{laflamme,bennett} onto an 8-qubit 
code~\cite{gottesman,calderbank2,steane2}.  Since the 5-qubit code has four 
generators, while the 8-qubit code has only five, we must first augment the 
8-qubit code by adding a trivial sixth generator.  The resulting stabilizer 
(using the stabilizer from \cite{gottesman} for the 8-qubit code and from 
\cite{calderbank2} for the 5-qubit code) is given in table~\ref{code13}.  
Since the stabilizer has six generators, this code encodes seven qubits in 13 
qubits.  This is the best code on 13 qubits allowed by the quantum 
Hamming bound.
\begin{table}
\begin{tabular}{|l|ccccccccccccc|}
$M_1$ & $X_1$ & $X_2$ & $X_3$ & $X_4$ & $X_5$ & $X_6$ & $X_7$ & 
$X_8$ & $ I $ & $ I $ & $ I $ & $ I $ & $ I $\\
$M_2$ & $Z_1$ & $Z_2$ & $Z_3$ & $Z_4$ & $Z_5$ & $Z_6$ & $Z_7$ & $Z_8$ 
& $ I $ & $ I $ & $ I $ & $ I $ & $ I $\\
$M_3$ & $X_1$ & $ I $ & $X_3$ & $ I $ & $Z_5$ & $Y_6$ & $Z_7$ & $Y_8$ 
& $X_9$ & $X_{10}$ & $Z_{11}$ & $ I $ & $Z_{13}$\\
$M_4$ & $X_1$ & $ I $ & $Y_3$ & $Z_4$ & $X_5$ & $ I $ & $Y_7$ & $Z_8$ 
& $Z_9$ & $X_{10}$ & $X_{11}$ & $Z_{12}$ & $ I $\\
$M_5$ & $X_1$ & $Z_2$ & $ I $ & $Y_4$ & $ I $ & $Y_6$ & $X_7$ & $Z_8$ 
& $ I $ & $Z_{10}$ & $X_{11}$ & $X_{12}$ & $Z_{13}$\\
$M_6$ & $ I $ & $ I $ & $ I $ & $ I $ & $ I $ & $ I $ & $ I $ & $ I $ & $Z_9$ 
& $ I $ & $Z_{11}$ & $X_{12}$ & $X_{13}$\\
\end{tabular}
\caption{The stabilizer for $n=13$ formed by pasting an $n=5$ code to an 
$n=8$ code.}
\label{code13}
\end{table}
 
We can also paste a 5-qubit code to the 16-qubit code of the class given in 
\cite{gottesman}.  Since the 16-qubit code already has six generators, no 
augmentation is needed.  This produces a 21-qubit code encoding 15 
qubits.  This is the second perfect code.  In general, if we paste the 
$(j-1)$th perfect code (with $n=(4^j-1)/3$ and $2j$ generators) to a 
$n=2^{2j}$ code, we get a code with $2j+2$ generators on $4^j + (4^j-1)/3 = 
(4^{j+1}-1)/3$ qubits.  This is therefore the $j$th perfect code, and we can 
produce all the perfect codes using this construction.

Pasting other combinations of codes is also possible, but not all 
combinations can be used.  The larger code must always have a generator 
formed from the product of all $X_i$s and a generator equal to the product 
of all $Z_i$s, or some equivalent set of generators that can be used to 
distinguish errors on the original set of qubits from those on the new 
qubits added after the pasting operation.  Both codes must be nondegenerate,
because when $F^\dagger E$ is in ${\cal H}$ before the pasting, it is unlikely
to remain in ${\cal H}$ after the pasting operation, which lengthens most 
of the generators.  Also, the smaller code must have 
exactly two fewer generators than the larger code.  This requirement can 
be largely circumvented, however, by adding on identity generators to 
either the larger or smaller code, as seen in the above construction of a 
13-qubit code.

In addition, this method does not work at all on codes to correct two or 
more errors. Suppose we used a similar method to 
distinguish one- or two-qubit errors on the original qubits from those on 
the new qubits. A new two-qubit error formed of one 
error on the original qubits and one on the new qubits would look like 
an error on the original qubits, since it does actually affect them, and
would not typically be distinguishable from errors on the original qubits.
Of course, in some special cases, the new code might distinguish such
errors, but we cannot be sure that it will based purely on the pasting
method described here.

\section*{Acknowledgements}

I would like to thank John Preskill for helpful comments, as well as the
ISI Foundation for sponsoring the Quantum Computation workshop, where much
of this work was performed.  This work was supported in part by the U.S.
Department of Energy under Grant No. DE-FG03-92-ER40701 and by DARPA
through a grant to ARO.

\end{document}